\documentclass[aps,prb,twocolumn,showpacs,floatfix]{revtex4-1}
\usepackage{graphicx,color}
\usepackage{amsfonts}
\usepackage[figuresright]{rotating}  
\usepackage{amssymb}
\usepackage{amsmath}
\usepackage{mathtools}
\usepackage{psfrag}
\usepackage{subfigure}
\usepackage{multirow}
\usepackage{tabularx}
\usepackage{textcomp}
\usepackage{units}
\usepackage{hyperref}

\def\nn{\nonumber}

\def\beq{\begin{eqnarray}}
\def\eeq{\end{eqnarray}}

\renewcommand{\v}[1]{\ensuremath{\mathbf{#1}}} 
 
\let\baraccent=\= 
\renewcommand{\=}[1]{\stackrel{#1}{=}} 

\makeatletter

\begin{document}
\title{Dynamics of tunneling into nonequilibrium edge states}
\author{Benjamin M.\ \surname{Fregoso}}

\affiliation{Department of Physics, University of California, Berkeley, CA, 94720, USA} 

\author{Jan P.\ \surname{Dahlhaus}}

\affiliation{Department of Physics, University of California, Berkeley,  CA, 94720, USA} 

\author{Joel E.\ \surname{Moore}}

\affiliation{Department of Physics, University of California, Berkeley,  CA, 94720, USA} 

\begin{abstract}
Time-dependent perturbations can drive a trivial two-dimensional band insulator into a quantum
Hall-like phase, with protected nonequilibrium states bound to its edges.
We propose an experiment to probe the existence of these topological edge states which consists of passing a 
tunneling current through a small two-dimensional sample out of equilibrium. 
The signature is a nonquantized metallic conductance near the edges of the sample and, in contrast, an excitation gap in the bulk.
This proposal is demonstrated for the case of a
two-dimensional lattice model of Dirac electrons with tunable mass in a strong electromagnetic field. 
In addition, we also study the tunneling conductance of the 
driven resonant level model and find a phenomenon similar to  
dynamic localization in which certain transport channels are suppressed.
\end{abstract}

\maketitle

\section{Introduction}
A key signature of topological matter is the existence of protected edge states. 
The advent of topological insulators\cite{Hasan2010,kane&mele-2005,Bernevig2006,Moore2007,fu&kane&mele-2007,Roy2009} 
which exhibit such states protected by time-reversal symmetry, has caused a great interest
in these states, both for fundamental reasons and because of their potential applications.
Recently, it has been proposed to engineer topological band structures
in nonequilibrium matter\cite{Lindner2011} where an initially topologically trivial insulator is converted into a 
topological insulator via an external time-dependent perturbation. 
For a periodic perturbation, these nonequilibrium states could be characterized in certain regimes 
by the corresponding Floquet Hamiltonian\cite{Shirley1965}
which can have a non-trivial topology\cite{Rudner2013}.   
Examples of the richness of time-domain phenomena in the solid state  
can be found in Refs.~\onlinecite{Kadanoff1989,Tien1963,Jauho1994,Keay1995,Wang2013,Onishi,Fregoso2013,Sentef,Shen,Perez-Piskunow2014,Gomez-Leon2013a}.

Even though Floquet bands have been observed in a solid state context recently\cite{Wang2013,Onishi,Fregoso2013}, 
there is no evidence of any topological aspect associated with them yet. 
The gaps observed so far are not topological in nature and are
better understood as avoiding crossings of bands of effective
Hamiltonians, see Ref.~\onlinecite{Fregoso2013} for details. Indeed, they arise
as pure electric field effects and should not be interpreted
literally as a signature of time reversal symmetry breaking.
The aim of this work is (i) to propose an experimental setup where the 
nonequilibrium edge states could be realized, and
(ii) if edge sates are realized, to unambiguously characterize them by their tunneling spectra.
As byproduct, we obtain the complete analytic tunneling current spectra and occupation of a driven quantum dot
with one and two levels.

Scanning Tunneling Microscopy (STM) has been a useful tool in understanding electronic properties of condensed matter
systems\cite{Bai2000}. Electron tunneling provided the earliest evidence of an electronic excitation gap in conventional superconductors\cite{Giaever1960}.
The spatial resolution of STM has played a key role in understanding the chiral edge states in graphene nanoribbons
with different edge geometries\cite{Tao2011}.
As will be explained bellow, the advantage of STM is that it allows for spatial 
resolution which unequivocally differentiates bulk from edge states and, crucially, 
maps out the time-average of the density of current-carrying states of the two-dimensional sample, 
hence it is not necessary that the edge states are occupied, all that is required in our setup is that
they carry finite spectral weight.

\begin{figure}[t]
\subfigure{\includegraphics[width=0.40\textwidth]{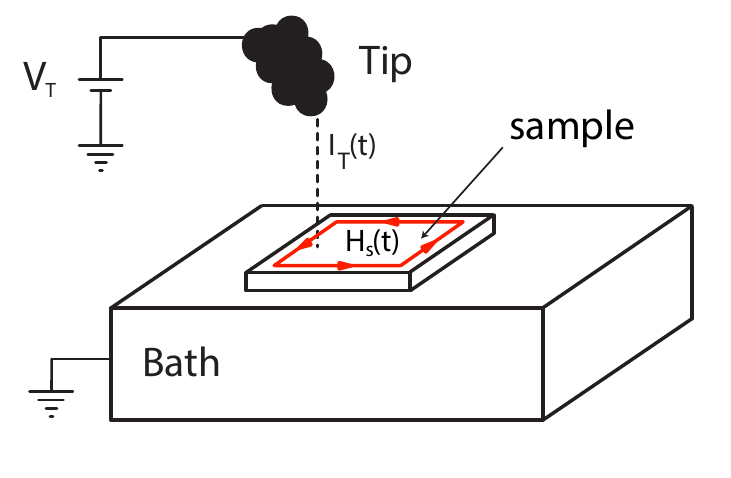}}
\caption{(Color online) STM setup for tunneling into nonequilibrium states. The Hamiltonian of 
a two-dimensional small sample is time-dependent. The fermionic reservoirs are always in equilibrium. 
We are interested in the changes to the tunneling current at the location of the tip due to the time-dependent perturbation.
In particular, we aim to probe nonequilibrium edge states (indicated by arrows) that arise due to the time dependence.}
 \label{fig:stmsetup}
\end{figure}
To illustrate how tunneling spectroscopy translates to driven systems, we first consider the tunneling
spectra of a driven quantum dot which is analytically solvable. In the long-time limit, tunneling into the dot
state gives non-quantized peaks in the conductance at energies corresponding to the Floquet states. Incidentally,
we find that for certain driving amplitudes the tunneling conductance vanishes, 
a phenomenon also seen in driven semiconductors\cite{Keay1995}.

Next we study tunneling into a two-dimensional lattice model driven by a 
circularly polarized electromagnetic field (EM). Here we consider the coupling to the reservoirs, 
the effect of finite temperature and a specific switch-on time of the field. 
In the absence of driving, the conductance spectrum shows a gap
across the sample, indicating insulating behavior. However, when the EM field is turned on, a new 
(Floquet) band structure forms after a transient period.  More precisely, once the wave envelope is constant
so the perturbation is strictly periodic, the Floquet states become quasi-stationary states, i.e., they are stationary
with respect to advances in time by multiples of the drive period.  In the long-time dynamics, the conductance is nonzero
at the edges of the sample, indicating the presence of localized Floquet states. The bulk gap persists. 
The magnitude of the conductance depends on the nonequilibrium spectral weight carried by these states at the chemical
potential of the tip. For small amplitude driving, it is mostly localized around the original band structure.
Our purpose here is to suggest a possible realization of nonequilibrium edge states with short
ultrafast pulsed-laser excitations where relaxation mechanisms play a small role,
and to understand the STM signatures of nonequilibrium edge states \text{if} 
the Floquet regime is achieved.

In Sec. \ref{sec:model_driven}, we describe the model Hamiltonian of the proposed STM setup. 
Then in Sec. \ref{eq:tunneling_curr} we use nonequilibrium perturbation theory to compute the tunneling current.  
In Sec. \ref{sec:driven_one_level}, we study the tunneling spectrum and occupation of a driven resonant level. 
In Sec. \ref{sec:num_discuss} we discuss the model of nonequilibrium Dirac Fermions and conclude in \ref{sec:conclusions}.
The Appendixes explain the details of the calculations.

\section{STM model Hamiltonian} 
\label{sec:model_driven}
We consider a two-dimensional sample as shown in Fig.~\ref{fig:stmsetup} which
is subject to a time-dependent driving field. A moveable STM tip is located close above
the sample, tunnel-coupled locally to the states in the system. The sample is further connected weakly to a
bath located beneath a layer of insulating substrate. By choosing one local and one global probe, the experiment
can gather local information by moving just a single tip. The system can be described by a generic model with
explicit time dependence,

\begin{align}
\hat{H}(t)=\hat{H}_S(t) + \hat{H}_{\alpha} + \hat{H}'
\label{eqn:full_H}
\end{align}
where,
\begin{align}
\hat{H}_{S}(t) &= \sum_{is j\sigma} H_{isj\sigma}(t) c^{\dagger}_{is}  c_{j\sigma}  &\nonumber \\
& \hspace{35pt}+ \frac{1}{2}\sum_{} c^{\dagger}_{is} c^{\dagger}_{js} U_{is,js',k\sigma,l\sigma'} c_{l\sigma'} c_{k\sigma}, &\\
\hat{H}_{\alpha} &= \sum_{k s} \epsilon_{\alpha k} a^{\dagger}_{\alpha k s} a_{\alpha k s}, & \\
\hat{H}' &= \sum_{\alpha k\sigma,is}  J_{ \alpha k\sigma,is} a^{\dagger}_{\alpha k\sigma} c_{is}  + h.c.,&  
\end{align}
and $H_S(t)$ is the sample Hamiltonian of interest, $H_\alpha$ are the tip ($\alpha=T$) and bath 
($\alpha=B$) fermionic reservoirs. Both leads are considered large unpolarized, nonsuperconducting Fermi liquids. 
$H'$ is the sample coupling to the tip and bath. The tip couples  
to a particular site $i=T$ of the sample, 
$J_{T k\sigma ,is}=\delta_{\sigma s}\delta_{i T} J_{Tk s,Ts}$.
We also use this symbol for temperature and also to distinguish functions and variables of
the tip (like chemical potential $\mu_T$) from those of bath (e.g. $\mu_B$).
We hope the meaning will be clear from the context. For simplicity, the  
 bath is assumed to have no spatial structure and couples to all sites such that 
$J_{B k\sigma ,is}=\delta_{\sigma s} J_{B k s,i s}$.  
Indices such as $\sigma,s,s'$ label the particle spin and $i,j$ 
represent a set of atomic levels at each site, for example $i_s, i_{px}, i_{py}...$ would correspond to an $s$ and $p$ levels
on site $i$. $k$ labels the three-dimensional (3D) momentum of the bath and tip degrees of freedom. 
The sample dynamics is represented by a \textit{time-dependent} 
tight-binding Hamiltonian with possible interactions represented by $U_{is,js',k\sigma,l\sigma'}$.

\section{tunneling current}
\label{eq:tunneling_curr}
Here we follow  Ref.~\onlinecite{Jauho1994} and rewrite the transport problem in 
the language of correlators in the Baym-Kadanoff contour\cite{Kadanoff1989}. 
The current through the tip is 
\begin{align}
I_T(t) = -e \langle d N_T(t)/dt \rangle.
\end{align}
The operators are in the Heisenberg representation with respect to $\hat{H}(t)$ and 
$\langle \cdot \rangle = \textrm{Tr}(\rho_0\cdot)$ is the average over initial density 
matrices of the total system including reservoirs. 
The number of particles in the tip is $N_T(t)=\sum_{k s} a^{\dagger}_{T k s}(t) a_{T k s}(t)$ and
 $d N_T/d t = i[N_T,\hat{H}(t)]$ can be expressed as an equal-time mixed correlator  
$G^{<}_{is,\alpha k s'}(t,t') \equiv i \langle a^{\dagger}_{\alpha k s'}(t') c_{is}(t) \rangle$,
\begin{align}
I_T(t) &= \frac{2 i e}{\hbar}\textrm{Re}\sum_{k\sigma s}  J_{T k\sigma ,Ts} \langle a^{\dagger}_{T k\sigma}(t) c_{Ts}(t) \rangle \nonumber \\
&=\frac{2 e}{\hbar}\textrm{Re}\sum_{k\sigma s}  J_{T k\sigma ,Ts} G^{<}_{Ts,Tk\sigma}(t,t).
\label{eqn:gen_current1}
\end{align}
The tunneling Hamiltonian $H'$ is connected to the sample in the far past where each 
component of the system is separately in equilibrium. The tip and the bath initial density matrix are defined by their 
temperature and chemical potential and are assumed to be in equilibrium at all times.
After a transient period the sample reaches a nonequilibrium steady state which is 
determined by the state of the reservoirs.
We perform perturbation theory of
$G^{<}_{Ts,Tk\sigma}(\tau,\tau') = \langle T_c a^{\dagger}_{T k\sigma}(\tau) c_{Ts}(\tau') \rangle$ 
on the closed time path contour\cite{Kadanoff1989} 
with respect to the tunneling term and obtain an expression in which the reservoir correlators are factored out from the 
sample correlators. In the final step contour ordered correlators are projected into 
real time\cite{Langreth1991}, 
\begin{align}
G^{<}_{is,\alpha k\sigma}(t,t')&= \sum_{js'}\int dt_1 J^{*}_{\alpha k\sigma,js'} [G^{r}_{is,js'}(t,t_1)g^{<}_{\alpha k \sigma}(t_1,t') \nn \\
&+ G^{<}_{is,js'}(t,t_1) g^{a}_{\alpha k\sigma}(t_1,t')],
\label{eqn:G_less}
\end{align} 
where,  
$G^{r,a}_{is,js'}(t,t') \equiv- i\theta(\pm t \mp t')\langle\{ c_{js'}(t), c^{\dagger}_{is}(t')\} \rangle$,
$G^{<}_{is,js'}(t,t') \equiv i \langle c^{\dagger}_{is}(t') c_{js'}(t)  \rangle$,
$g^{r,a}_{\alpha k \sigma}(t,t') \equiv- i\theta(\pm t \mp t')\langle \{a_{\alpha k \sigma}(t), a^{\dagger}_{\alpha k \sigma}(t') \}\rangle$,
$g^{<}_{\alpha k \sigma}(t,t') \equiv i \langle  a^{\dagger}_{\alpha k \sigma}(t') a_{\alpha k \sigma}(t) \rangle$.
Crucially, we switch on interactions in the far past so that the actual 
time dependence of observables is only due to an external time-dependence of the Hamiltonian with no memory of initial state\cite{Kadanoff1989}. 
The tip and bath operators are assumed non-interacting and hence $g^{r,a}_{\alpha k \sigma}(t,t') = \mp \theta(\pm t \mp t') \exp[-i \epsilon_{\alpha k}(t-t')]$
and $g^{<}_{\alpha k \sigma}(t,t') = i f_{\alpha}(\epsilon_{\alpha k}) \exp[-i \epsilon_{\alpha k}(t-t')]$,
are found from their equations of motion. But the operators in $G^{r,<}(t,t')$ are in the interaction representation 
with respect to $\hat{H}_S(t)$ and hence must be computed taking into account the coupling to the reservoirs. Explicitly, from 
Eqn.~\ref{eqn:gen_current1} we obtain\cite{Jauho1994},
\begin{align}
I_{T}(t) &= -\frac{2e}{\hbar} \textrm{Im}\hspace{-3pt}\int_{{-\infty}}^{t}\hspace{-8pt} dt_1 \hspace{-3pt}\int\hspace{-3pt} \frac{d\epsilon}{2\pi} e^{-i \epsilon(t_1-t)} \Gamma_{T}(\epsilon)[ G_{TsTs}^{<}(t,t_1) \nn \\
&\hspace{90pt}+ f_{T}(\epsilon)G_{TsTs}^{r}(t,t_1)].
\label{eqn:gen_current}
\end{align}
Summation over repeated spin indices is implied and we have defined
$\Gamma^{T}_{is,js'}(\epsilon) = 2\pi \sum_{\sigma} \rho_{T\sigma}(\epsilon) J_{T\sigma,is}(\epsilon) J^{*}_{T\sigma,js'}(\epsilon) 
= \delta_{ss'}\delta_{i,T}\delta_{j,T} 2\pi \sum_{\sigma} \rho_{\sigma}(\epsilon) | J_{T\sigma,T\sigma}(\epsilon)|^2 = \delta_{ss'}\delta_{i,T}\delta_{j,T} \Gamma_{T}(\epsilon)$ and
 $J_{T k\sigma,is} = J_{T,\sigma,is}(\epsilon)$. The Fermi distribution $f_{T}(\epsilon)=(1+\exp{\beta(\epsilon-\mu_T)})^{-1}$ describes the thermal occupation of the tip.
In Eqn.~\ref{eqn:gen_current} all correlation and nonlinear effects are included in the sample local Green functions.
For the rest of this work we apply Eqn.~\ref{eqn:gen_current} 
to the problem of tunneling into driven non-interacting nonequilibrium states.

\subsection{Time-dependent non-interacting states}
\label{sec:tunneling}
The tunneling into and and out of the sample is taken into account to all orders of perturbation theory via a complex self-energy
in the Dyson equations\cite{Kamenev2011} 
\begin{align}
\v{G}^{r,a}&=\v{G}_0^{r,a} + \v{G}_0^{r,a}\v{\Sigma}^{r,a} \v{G}^{r,a},  \label{eqn:gr_ret}\\
\v{G}^{<}&= \v{G}^{r}\v{\Sigma}^{<} \v{G}^{a}.    \label{eqn:gr_gless}
\end{align}
Time variables have been omitted for clarity and integration  
of repeated ones over the real axis is implied. The self-energies are
\begin{align}
\Sigma^{r,a}_{is,js'}(t,t') &= \sum_{\alpha k \sigma} J^{*}_{\alpha k \sigma,is} g^{r,a}_{\alpha k\sigma}(t,t') J_{\alpha k \sigma,js'},  \label{eqn:sigma_ret_adv}\\
\Sigma^{<}_{is,js'}(t,t') &=i \sum_{\alpha} \hspace{-3pt} \int \hspace{-3pt}\frac{d\epsilon}{2\pi} f_{\alpha}(\epsilon)\Gamma_{is,js'}^{\alpha} e^{-i\epsilon(t- t')},
\label{eqn:dyson}
\end{align}
where we assumed reservoirs with featureless density of states, i.e., 
the wideband limit (WBL), where   
$J_{T k\sigma ,is}=\delta_{\sigma s}\delta_{i T} J_{T}$ and $J_{B k\sigma ,is}=\delta_{\sigma s} J_{B}$
(see Appendix \ref{sec:wbl}).
$\Gamma^{T}_{is,js'} = 2\pi \sum_{\sigma} \rho_{T\sigma} J_{T\sigma,is} J^{*}_{T\sigma,js'}$
is the tip-sample coupling.
From Eqn.~\ref{eqn:sigma_ret_adv} we see that the retarded self energy is Markovian  
$\Sigma^{r,a}_{isjs'}(t,t') =\mp (i/2)(\Gamma^{T}_{isjs'} + \Gamma^{B}_{isjs'})\delta(t-t')=\mp (i/2)\delta_{ss'}(\delta_{i,T} \delta_{j,T}\Gamma_T + \Gamma_{B})\delta(t-t')$,
 and hence Eqn.~\ref{eqn:gr_ret} adopts the differential form, $(i \partial_t - H_S(t)- \Sigma^r)G^{r}(t,t') = \delta(t-t')$ with  
boundary conditions $G^{r}(t,t')=0$ for $t < t'$, $G^{r}(t,t)=-i$. 
In general, $\Sigma^{r}$ and $H_S(t)$ do not commute 
and we must Trotterize the time evolution of $G^{r}(t,t')$. Once the retarded Green function is known
all other correlators can be obtained from it using Eqn.~\ref{eqn:gr_gless}. Some analytic progress can be made 
for strictly periodic Hamiltonians (Appendix \ref{sec:periodic_H}) and
the driven one- and two-level systems discussed in the next section.

\section{Tunneling into a nonequilibrium quantum dot}
\label{sec:driven_one_level}
\subsection{Quantum dot with one level}
\begin{figure}[t]
\subfigure{\includegraphics[width=0.45\textwidth]{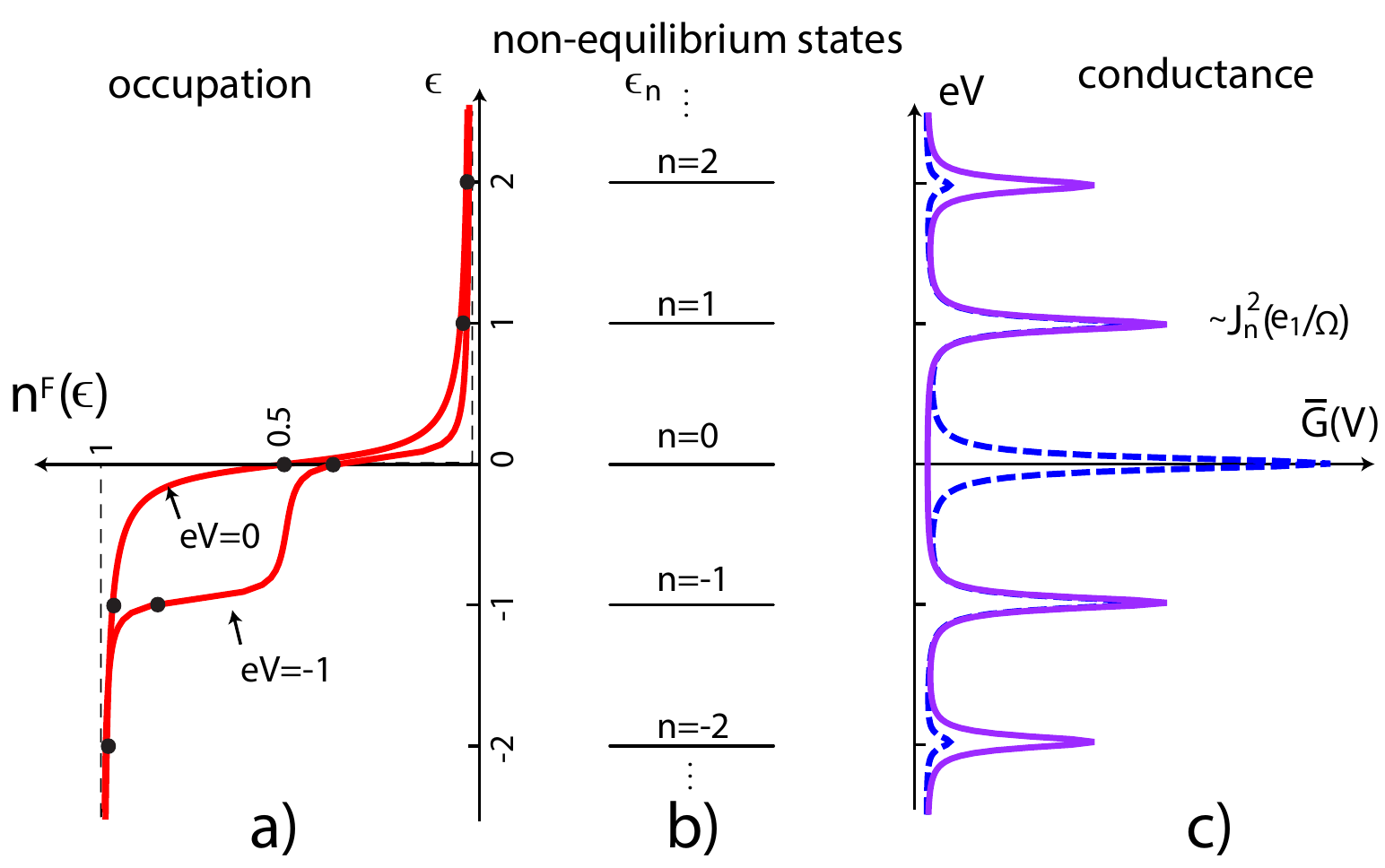}}
\caption{(Color online) (a) Occupation of nonequilibrium states
in the driven quantum dot with one level; see  Eqn.~\ref{eqn:occ_one_level}. Here we chose representative values of 
the bias eV$=0,-1$ and $\bar{\Gamma}= 0.1$ (solid curves). The dashed curve corresponds to eV$=0=\bar{\Gamma}$.
(b) Nonequilibrium states of a driven quantum dot with one level. 
(c) Tunneling spectra [Eqn.~\ref{eqn:cond_one_level_zero_T}] for two amplitudes of the drive $e_1=1.2, 2.4$. For 
$e_1 = 2.4$(solid) we have $J^{2}_0(e_1/\Omega)=0$ and the zero-bias resonance vanishes. In this case, weight is 
transferred to higher order nonequilibrium states. We set $\mu_B =0=e_0=T$ in all panels and energies are in units of $\Omega$.}
 \label{fig:occ_cond_one_level}
\end{figure}
It is useful to consider the case in which a complete analytical solution exists, i.e., that of a
periodically driven one level connected to featureless fermionic reservoirs. 
It serves to illustrate the main features in the tunneling
current through a driven system\cite{Tien1963} and translates to a large
degree to the more complicated situation of a two-dimensional sample discussed in the next section.
Incidentally, we also show an example where 
the conductance vanishes for certain values of the driving amplitude, a 
dynamic localization\cite{Keay1995} of sorts where the electron is prevented from tunneling due to 
quantum interference effects. Let the Hamiltonian (in first quantization) be 
\begin{align}
H_{d}(t) = e_{0} + e_1 \cos \Omega t. 
\end{align}
For this model, the quasienergy is $e_0$ and the Floquet states are labeled as $\epsilon_n=e_0 + n\Omega$. 
The Dyson equation takes a scalar form
$(i\partial_t - e_0 - e_1 \cos\Omega t + i\bar{\Gamma})G^{r}(t,t') = \delta(t-t')$ where 
$\bar{\Gamma} = (\Gamma_T + \Gamma_B)/2$. 
In the WBL Eqn.~\ref{eqn:gen_current} gives the time-averaged current,
\begin{align}
\bar{I}_T &= \frac{e}{\hbar}~\frac{\Gamma_T  \Gamma_B}{\Gamma_T+\Gamma_B}  \int d\epsilon  ~\rho_{d}(\epsilon)[f_{T}(\epsilon)- f_{B}(\epsilon) ],
\label{eqn:curr_one_level_landauer}
\end{align}
where the time-averaged density of states (see Appendix \ref{sec:periodic_H}),
$\rho_d(\epsilon)= -(1/\pi)\textrm{Im} G^{r}(0,\epsilon)=(\bar{\Gamma} /\pi)\sum_n J_{n}^2(e_1/\Omega)/((\epsilon - \epsilon_n)^2 + \bar{\Gamma}^2)$ 
has resonance peaks at the eigenvalues of the Floquet Hamiltonian and 
satisfies the sum rule $\int d\epsilon \rho_d (\epsilon)=1$ due to fermion conservation\cite{Fregoso2013}.
$J_n$ in the nth Bessel function of the 1st kind. Intuitively, the state's original spectral weight breaks 
into infinitely many `states' labeled by $n$ each broadened by the coupling to
the reservoirs. Integration over energy gives
\begin{align}
\bar{I}_T &= \frac{e}{h}\frac{\Gamma_T  \Gamma_B}{\bar{\Gamma}} \sum_n J_n^2(e_1/\Omega)\textrm{Im}[\psi_T^n -\psi_B^n ],
\label{eqn:avg_current_one_level}
\end{align}
where $\psi_{\alpha}^{n}=\psi(1/2 + ( \mu_\alpha-\epsilon_n  - i\bar{\Gamma})(i \beta/2\pi))$ and
$\psi(z)$ is the digamma function at $z$.  
This expression shows that tunneling electrons can gain $n$ photons of energy in the process and
that all n-th order processes contribute. Without loss of generality let 
$\mu_B=0$ and $\mu_T=eV$ then the time-averaged conductance $d \bar{I}_T/dV$ is  
\begin{align}
\bar{G}(V)&=\frac{e^2}{h} \frac{\beta}{2\pi} \frac{\Gamma_T  \Gamma_B }{\bar{\Gamma}} \sum_n J^2_{n}(e_1/\Omega)\times \nn \\
 &\hspace{25pt} \textrm{Re} ~\psi'(1/2 + ( eV-\epsilon_n - i\bar{\Gamma})(i \beta/2\pi)),
\label{eqn:cond_one_level_finite_T}
\end{align}
which at zero temperature becomes
\begin{align}
\bar{G}(V)=\frac{e^2}{h} \sum_n J^2_n(e_1/\Omega)\frac{ \Gamma_T\Gamma_B }{(eV-\epsilon_n)^2+\bar{\Gamma}^2}.
\label{eqn:cond_one_level_zero_T}
\end{align}
The important point is that the conductance has resonance peaks at 
energies where the nonequilibrium states have nonvanishing weight regardless of their occupation. 
The conductance is, in general, not quantized. Because the dot has no spatial structure,  
the tunneling conductance is simply related to the time-averaged density of states.
This feature remains in the driven lattice model in the tunneling regime.  The same sum of squares of Bessel functions appears
in a related quantity, the lifetime on the dot~\cite{grifoni}.

Interestingly, the tunneling resonance at $\epsilon_n$ vanishes for values of the driving amplitude  
which give zeroes of the Bessel functions. For example, if 
$e_1/\Omega = 2.41$ the resonance at $e_0$ vanishes [Fig.~\ref{fig:occ_cond_one_level}c].
Similarly, at $e_1/\Omega = 3.83$ the resonance at $e_0 +\Omega$ vanishes.
Since the total area under $\bar{G}(V)$ is conserved,
$\int d V \bar{G}(V) = (e/\hbar) \Gamma_T \Gamma_B/2\bar{\Gamma}$, other 
resonance peaks increase.  A similar phenomenon has been 
observed in driven semiconductors\cite{Keay1995} and it can  
give rise to negative conductance. 

We now consider the occupancy of the level,
$n_d(t)=-i G^{<}(t,t) = \langle c^{\dagger}(t) c(t)\rangle$. Since all the correlators 
are known, from Eqn.~\ref{eqn:gr_gless} we obtain the time-averaged occupation of the 
state as
\begin{align}
\bar{n}_d &= \sum_n J^2_n(e_1/\Omega) n^{F}(\epsilon_n),
\label{eqn:occ_one_level}
\end{align}
where we defined  
$n^{F}(\epsilon_n)  =1/2 + \textrm{Im} [\Gamma_T \psi^{n}_{T}+ \Gamma_B\psi^{n}_{B}]/2\pi\bar{\Gamma}$. 
This `Floquet occupation' is determined by the characteristics of the reservoirs and the coupling to them. 
For example, if $\mu_T=\mu_B=0$ and at zero temperature,  
$n^{F}(\epsilon) = \theta(-\epsilon)+ (1/\pi)\textrm{sign}(\epsilon) \arctan (\bar{\Gamma}/|\epsilon|)$
which has a similar structure as a Fermi distribution but with a second term due to the coupling to the 
reservoirs which introduces an effective broadening. 
Roughly, Eqn.~\ref{eqn:occ_one_level} means that Floquet states above either reservoir's chemical potential will be 
less occupied than those bellow them. In general, $n^{F}$ 
has a double-step structure for $\mu_T\neq \mu_B$; see Fig.~\ref{fig:occ_cond_one_level}a.
Several interesting cases at zero temperature can be analytically computed. We set $e_0=0$ for simplicity, 
but this could easily be relaxed:  (i) $\bar{n}_d=1/2$ 
for zero chemical potentials. This can be understood intuitively considering that `half the weight' of the state is 
below zero energy and is fully occupied. (ii) 
$\bar{n}_d=3/4$ for $\mu_T \gg \mu_B=0$ and  $\Gamma_T=\Gamma_B\to 0$. Here again half of the state is below zero energy 
and hence fully occupied. The other half of the state has positive energy and is half occupied due to equal coupling to 
reservoirs, hence $\bar{n}_d =1/2 + 1/4 =3/4$. (iii) $\bar{n}_d=1/4$ 
for negative $\mu_T \ll \mu_B=0$ and $\Gamma_T=\Gamma_B\to 0$.
Here, half of the state is half occupied at negative energies and empty at positive energies, 
hence $\bar{n}_d = 1/2 \times 1/2 =1/4$.

\subsection{Quantum dot with two levels}
We have seen in the previous section that driven systems show a rich spectrum of phenomena. 
This is even more so if more than one level is present. To illustrate 
this we consider the case of tunneling into a site with an internal two-level system (TLS), 
\begin{align}
H_{tls}(t) = -\frac{\lambda}{2}\sigma_x \cos\Omega t + \frac{\lambda}{2} \sigma_y \sin\Omega t - \frac{h}{2} \sigma_z. 
\end{align}
This Hamiltonian could model a spin-\nicefrac{1}{2} particle in a static magnetic field in the z-axis and a rotating 
magnetic field in the x-y plane.
We can go through similar procedure as outlined and obtain the average current, conductance and total occupation of the dot.
We find that the expressions in Eqns.~\ref{eqn:curr_one_level_landauer}-\ref{eqn:occ_one_level} 
hold with the replacements $\epsilon_n \to \epsilon_{\beta n}=\epsilon_\beta - n\Omega$ ($\beta=1,2$) since now there are two 
quasienergies: 
$\epsilon_{1,0} =\Omega_0 + \Omega/2$, $\epsilon_{2,0} =-\Omega_0 + \Omega/2$ 
with $\Omega_0 = \sqrt{\lambda^2 + (\Omega-h)^2}/2$,
and $\sum_n J_{n}^{2}\to \sum_{\beta n} B_{\beta n}$ where 
$B_{\beta n}= [\delta_{n,1} + \delta_{n,0} - \sin\kappa (\delta_{\beta,1} - \delta_{\beta,2} )(\delta_{n,0} - \delta_{n,1} )]/2$,
and  $\sin\kappa = (\Omega-h)/2\Omega_0$.
In particular, the time-average density of states still satisfies a sum rule $\int d\epsilon \rho_{tls}(\epsilon) =2$ 
Note that the quasienergies are not independent since $\epsilon_{1,0}+\epsilon_{2,0}=0$ mod $\Omega$. 
In this specific model only two weights are nonvanishing for each of the two quasienergies.

We comment on the important features of the tunneling spectra. For the 
undriven case $\lambda=0$ the tunneling conductance shows resonances at $\pm h/2$ as expected. 
When the system is driven $\lambda\neq 0$, the resonance peak at $h/2$ splits into two peaks,
symmetrically placed about $\Omega_0$, i.e., at $\epsilon_{1,0} = \Omega_0 + \Omega/2$
and $\epsilon_{1,1} = \Omega_0 - \Omega/2$, in general, with unequal heights. The peak at $-h/2$ splits in the same way.  
The energy gap between the side bands is $\Omega$. 
For the special point $h=\Omega$, all four peaks have equal weight, $B_{\beta n}=1/2$ and 
the gap between the quasienergies is the smallest given by $\lambda$.
At values of the frequency that satisfy $\lambda^2 = 2\Omega h - h^2$ only three peaks are observed.
As for the occupation of the dot the same general considerations apply as for the driven one-level model.

\section{Tunneling into nonequilibrium edge states}
\label{sec:num_discuss}
\subsection{System Hamiltonian}
\begin{figure}[t]
\subfigure{\includegraphics[width=0.47\textwidth]{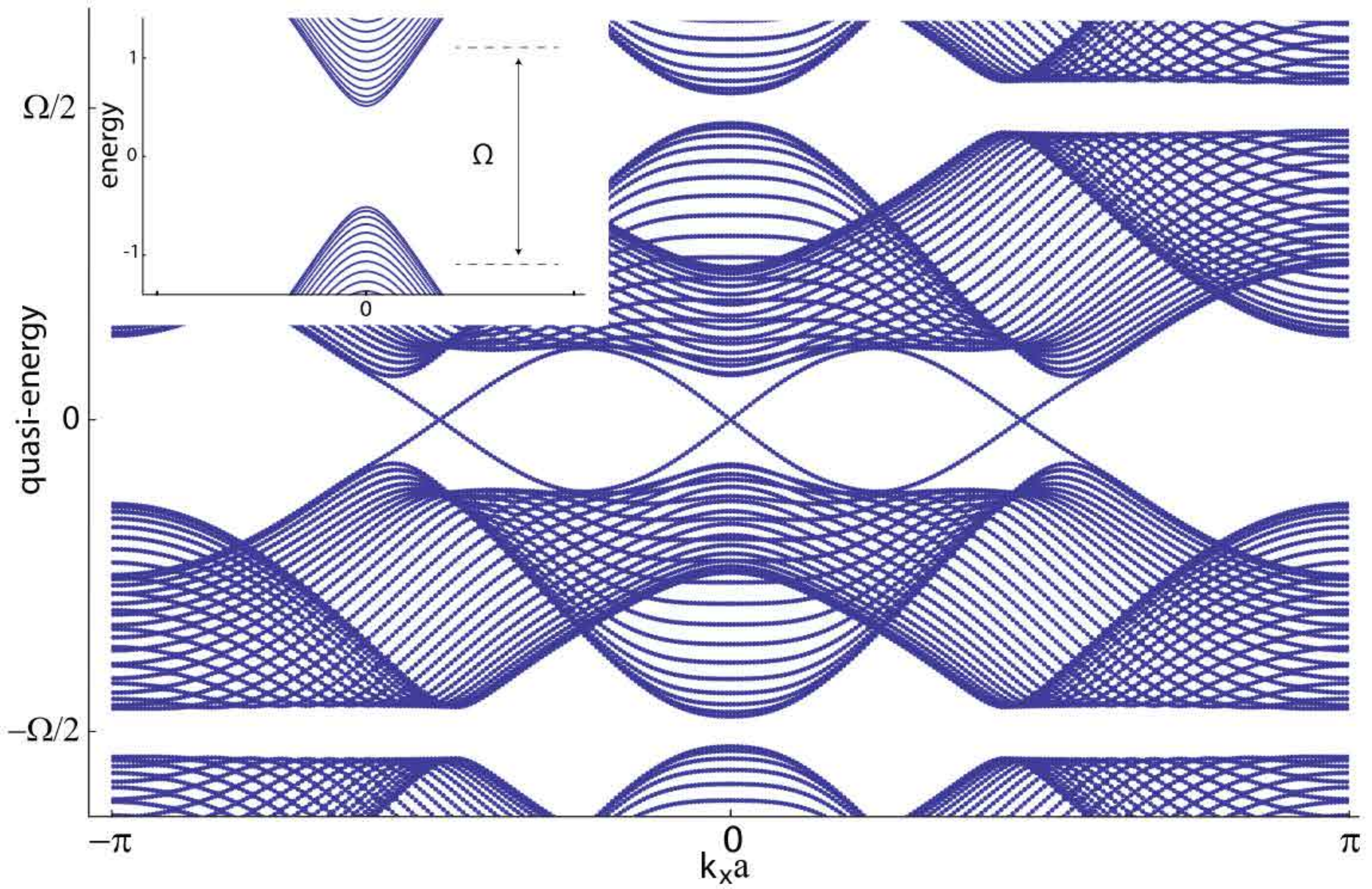}}
\caption{Typical quasienergy (and energy) spectrum in an isolated system on a ribbon 
geometry and with circularly polarized Electromagnetic fields. Since $\Omega$ is 
smaller than the bandwidth we see quasienergy band folding.
Six chiral edge states are visible, three of them propagate to one side and 
three to the other (at the opposite edges of the sample). Inset: band structure of the static Hamiltonian.
We truncate the Floquet Hamiltonian to $6$ modes and set $m=-2.5 A, e a \mathcal{A}_0/\hbar =1.0, \Omega =2.2 A, \Delta_{static}/A=1$.}
\label{fig:Floquet_spect}
\end{figure}

We now consider a model of non-interacting Dirac electrons with tunable mass on a square lattice strongly driven out of equilibrium. 
The static Hamiltonian in infinite space is 
$H(\v{k})=  A\sigma_x \sin k_x a   +  A\sigma_y \sin k_y a   + \sigma_z (m + A\cos k_x a  + A \cos k_y a )$.
The important energy scales are the gap $\Delta_{static}= 2|m+2|$ and the 
bandwidth max$\{m\pm 2A \}$. $a$ is the lattice constant and $m$ the mass of the Dirac particles.
Two copies of this model is realized in CdTe-HgTe quantum wells\cite{Bernevig2006}.
When $|m|<2A$ the bands have Chern number $\pm 1$ and hence 
when the system is put in a finite geometry, edge states(static) appear. Below we consider the 
case $|m|>2A$ on a finite square lattice.  The Hamiltonian describes electrons coupled to an EM field,  
\begin{align}
\hat{H}_{S}&(t) = \sum_{i} [ c^{\dagger}_i \frac{(i A \sigma_x + A\sigma_z)}{2} e^{-i e a \mathcal{A}_{x}(t)/\hbar} c_{i+\v{x}} + 
\label{eqn:edge_ham} \\
& c^{\dagger}_i \frac{(i A \sigma_y +  A\sigma_z)}{2} e^{-i e a \mathcal{A}_{y}(t)/\hbar} c_{i+\v{y}} + h.c. +  m c^{\dagger}_{i} \sigma_z c_{i} ],\nn
\end{align}
via minimal coupling where the hoppings acquire phases, 
$\exp[-i e a\v{\mathcal{A}}(t)\cdot (\v{R}_i-\v{R}_j)/\hbar]$, and $\v{R}_i$ is the 
position of the site $i$.
For circular polarization $\v{\mathcal{A}}(t)=\theta(t-t_0)\mathcal{A}_0 (\cos\Omega t,\sin\Omega t)$ where $\mathcal{A}_0=E_m/\Omega$ and 
we use a gauge in which the electric field is given by $\v{E}(t) = -\partial_t \v{\mathcal{A}}(t)$. Throughout 
small magnetic effects are ignored and we work in the tunneling regime, i.e.,  $\Gamma_T\ll \Gamma_B \ll A$.
A linearly polarized EM field also produces edge states and their STM 
signatures are similar. $t_0$ is the time when the drive is turned on and 
$E_m$ is the amplitude of the electric field on the surface.

\subsection{Isolated system on a ribbon geometry}

First we investigate the presence of edge states in an isolated system in a 
ribbon geometry with perfect time-periodic EM field, i.e., 
the driving has been turned on in the far past.
The quasienergy spectrum is shown in Fig.~\ref{fig:Floquet_spect}.  
For these plots we choose $e a \mathcal{A}_0/\hbar =1.0$, $m/A=-2.5$ and $\Omega=2.2 A$ which place the system 
in the non-topological regime with no irradiation (inset Fig.~\ref{fig:Floquet_spect}).
When the driving has been turned on and the Floquet regime reached the system is in a 
non-trivial topological state as verified by counting the edge states 
that cross the dynamical gaps\cite{Rudner2013} in the quasienergy spectrum.
Alternatively, we calculated the Chern number associated with the bulk bands in an infinite system, 
\begin{align}
C^{F}=\frac{1}{4\pi}\hspace{-3pt}\int_{BZ}\hspace{-4pt} d^2 k~ \hat{d}_F(\v{k})\cdot (\partial_{k_x}\hat{d}_F(\v{k})\times \partial_{k_y}\hat{d}_F(\v{k})),
\end{align}
where $\hat{d}_F(\v{k})$ is defined by the Floquet Hamiltonian as 
$\hat{H}_F\equiv \sum_\v{k}c^{\dagger}_{\v{k}\alpha} c_{\v{k}\beta} \hat{d}_{F}(\v{k})\cdot \vec{\sigma}_{\alpha\beta}$.
We obtain $C^{F}=3$ which indicates a non trivial topological state. 
There are three edge states propagating in the same direction, on each edge.
Moreover, the system undergoes dynamical quantum phase transitions as a function of the amplitude of the drive 
where one of the dynamical gaps closes: 
for $0.2<e a\mathcal{A}_0/\hbar< 0.95$ we have, within our numerical precision, $C^{F}=1$, for $0.95<e a\mathcal{A}_0/\hbar< 1.6$ we have $C^{F}=3$ 
and for $1.6<e a\mathcal{A}_0/\hbar< 1.7$,  $C^{F}=-1$.
In our calculations, we find evidence of edge states for arbitrarily small field amplitudes, however 
the gaps are small and the numerical cost to compute $C^{F}$ is large.

\begin{figure}[t]
\subfigure{\includegraphics[width=0.42 \textwidth]{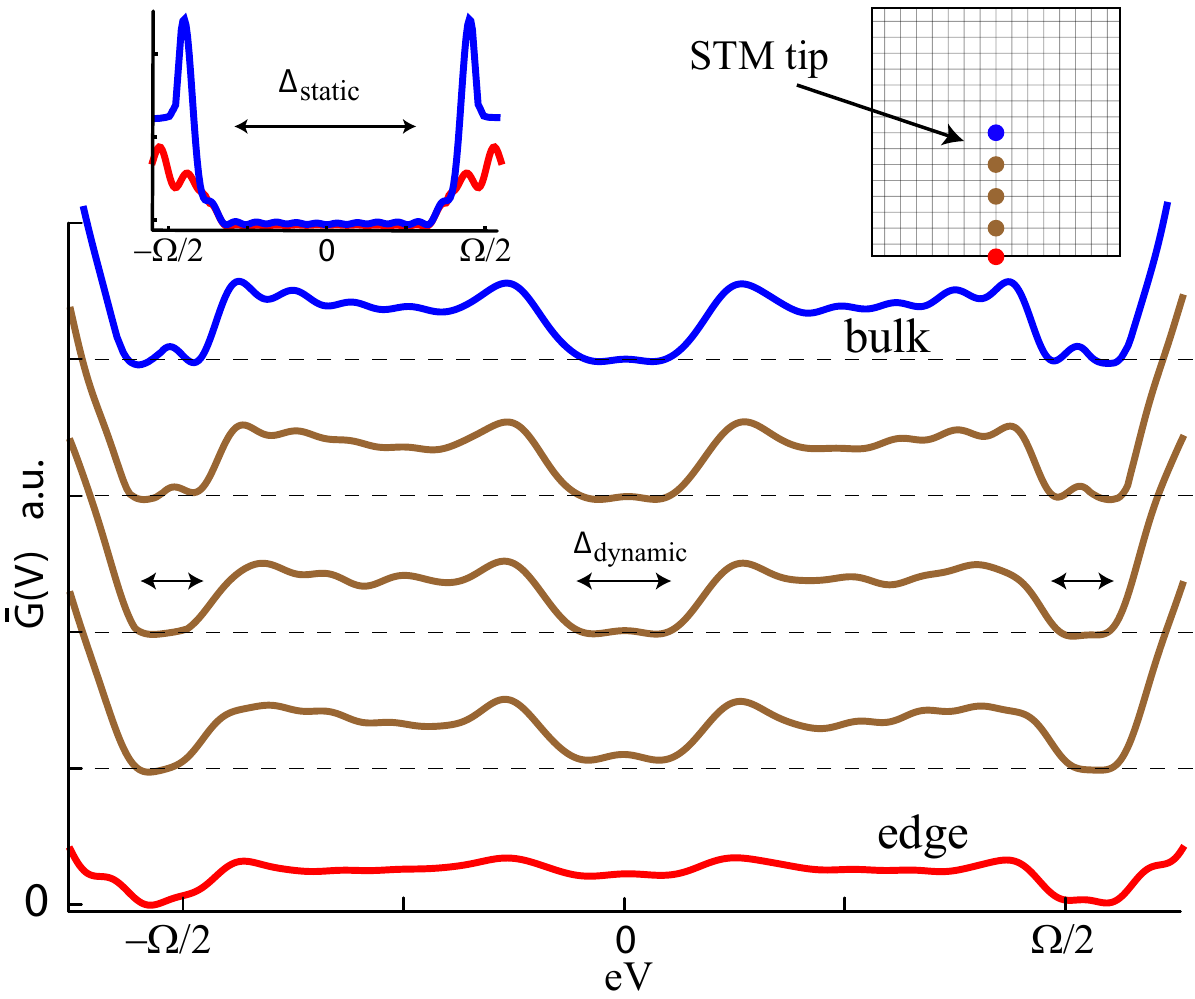}}
\caption{(Color online) Time-averaged tunneling conductance  
in a square geometry sample at different tip positions from bottom upward: (9,1)a, (9,3)a, (9,5)a, (9,7)a and (9,9)a; see 
right inset. The zero of each curve (dashed lines) was displaced vertically for clarity. The edges of the sample 
are gapless near zero energy while the bulk is gapped with two dynamical gaps visible: at zero bias and at the zone edges $\pm\Omega/2$.
The inset is the tunneling spectrum for the undriven system which shows an insulating behavior across the sample. We used a circularly polarized 
electromagnetic field with parameters as in Fig.~\ref{fig:Floquet_spect} and $T=0.001 A, \Gamma_B=0.01 A, \Gamma_T=0.005 A$,
$N_x\times N_y=17\times 17 = 289$ sites. Note that we work in the tunneling regime where $\Gamma_T \ll \Gamma_B \ll  A$.}
 \label{fig:conduc}
\end{figure}

\subsection{System on a square geometry connected to a bath and with switch-on time $t_0$}
The edge states remain present when going from a ribbon to a square geometry.
We chose a finite size sample of square shape with $N_x\times N_y=17\times 17 = 289$ sites.
Before the perturbation is turned on the sample is in a steady state determined by the reservoirs nearby. In the tunneling 
regime, the system is essentially unchanged from its time-independent state with zero chemical potential.
If the perturbation is turned on at a specific point in time $t_0$ we 
compute the conductance from  Eqn.~\ref{eqn:G_gen_time}. 
We find that it takes about 10-20 periods of the driving to reach the Floquet regime, see Appendix \ref{sec:wbl}.
Note that this considers only pure quantum evolution. In real driven macroscopic systems, it is not necessary 
that the system reaches a nonequilibrium steady state. If it does, the time elapsed will depend also  
on other additional microscopic relaxation mechanisms not considered here.

Fig.~\ref{fig:conduc} shows our numerical results for the time-averaged conductance in the Floquet regime. The STM tip 
is at various positions on the surface of the sample, e.g., in the center of the sample (blue) and at the edge 
(red). We first note that the time-averaged differential conductance vanishes in the bulk for 
energies lower than the dynamical gap that we found in the quasienergy band structure of Fig.~\ref{fig:Floquet_spect}. 
On the other hand, if the tip is located at the edge of the sample, the conductance at zero bias does not vanish 
indicating the presence of current carrying states there. This is a signature of the nonequilibrium edge states 
and one of our central results.
When the system is not driven, we see that the sample is insulating (top-left inset).
At bias voltages  approaching the quasienergies zone boundaries $\pm \Omega/2$ we observe dynamical gaps emerging 
in the STM signal as a depletion of current carrying states. If we had nonequilibrium edge states 
across this gap, the STM signal would be nonvanishing at $\pm \Omega/2$.
It is important to note that the STM spectrum is sensitive to the spectral weight of the nonequilibrium 
bands and hence is \textit{not} periodic in energy. We have verified that 
the small oscillations around zero conductance are due to finite size effects in addition to 
the sensitivity of the rapidly oscillation term in \ref{eqn:G_gen_time} at large bias voltages.  

To gain some intuition as to what it is probed by the conductance, let us assume 
the bath is connected only to the same site as the tip.  This gives a reasonable picture in the tunneling regime.
Then, one can show that the time-averaged tunneling conductance is a measure of the time-averaged nonequilibrium 
density of states. Indeed, Eqn.~\ref{eqn:curr_one_level_landauer} holds with (see Appendix \ref{sec:periodic_H})
\begin{align}
\bar{G}(V)\sim (e^2/\hbar)\Gamma_T \rho_{T}(V), 
\end{align}
where  $\rho_{T}(V) = -(1/\pi)\textrm{Im}\sum_s G^{r}_{TsTs}(0,V)$.
The role of the bath outside of the tunneling regime and with spatial extent is an interesting problem left for future 
investigations. In that case, the tunneling conductance is not a simple measure of the density of states 
but Eqn.~\ref{eqn:conductance_periodic} still holds. Note that in any case a vanishing conductance means 
the absence of nonequilibrium states. See also Ref.~\onlinecite{Perez-Piskunow2014} where 
transmission probabilities were calculated in a ribbon geometry for graphene.

\section{Discussion and conclusion}
\label{sec:conclusions}
We now turn to a discussion of the experimental relevance of our results. 
In a real experiment, decoherance effects such as phonons or particle 
interactions will make the situation more complex than considered above.
Specifically, these effects could modify the occupation of the nonequilibrium edge states.

As demonstrated in this work, the STM signal will probe the presence of current-carrying states, in particular 
Floquet modes(if present), irrespective of their occupation and their nature, i.e., if they are topologically trivial or not. 
However, we expect that the Floquet regime maybe achieved using ultra short pulsed laser excitations 
where relaxation processes have not fully taken place and the external drive 
provides the dominant dynamics. Such a regime has been achieved experimentally\cite{Wang2013}.
Indeed, our calculations are most relevant in the regime where the influence of  
phonons (or other decoherence effects) is weak, i.e., for time scales 
shorter than the phonon relaxation time, typically of order pico seconds.
Electron-electron interactions have relaxation times of the order of $10^{-15}$ seconds  
but this mechanism is less dominant in graphene and topological insulators. 
The interesting problem of the effect of these relaxation phenomena on the 
tunneling spectra of nonequilibrium edge states is left for a future study. 

While our calculations are based on a toy model, topological edge states are expected to occur in a 
variety of real systems upon radiation. The foremost interest has recently been on topological 
insulator surfaces and graphene. 
A pulsed laser field can reach electric field amplitudes 
$E_m= 2\times 10^7 V/m$ with a duration up to ps. We then we have a coupling  
$e a \mathcal{A}_0/\hbar\sim 0.02$, which is small. We note that if 
the chemical potential is close to the Dirac point, we also
expect light-induced gaps as, in that case, the important energy scale is given by
by \onlinecite{Fregoso2013} $e v E_m/\hbar\Omega^2$. These 
gaps, however, are not topological in nature. If a topological gap is achieved
of magnitude $\Delta \approx 50$ meV then a very crude estimate  
of the tunneling current from Eqn.~\ref{eqn:curr_one_level_landauer} gives  
$I\sim (e^2/h)\Gamma \rho V \sim 10^{-6} $ amps which appears for the duration of the pulse.

In conclusion, we have shown that an STM experiment can probe the existence of 
topological nonequilibrium edge states. In the process we further analyzed 
the tunneling dynamics into nonequilibrium states in a driven quantum dot. We found that
under certain driving conditions, transport channels can be suppressed
due to quantum interference effects.

\begin{figure}[t]
\subfigure{\includegraphics[width=0.40 \textwidth]{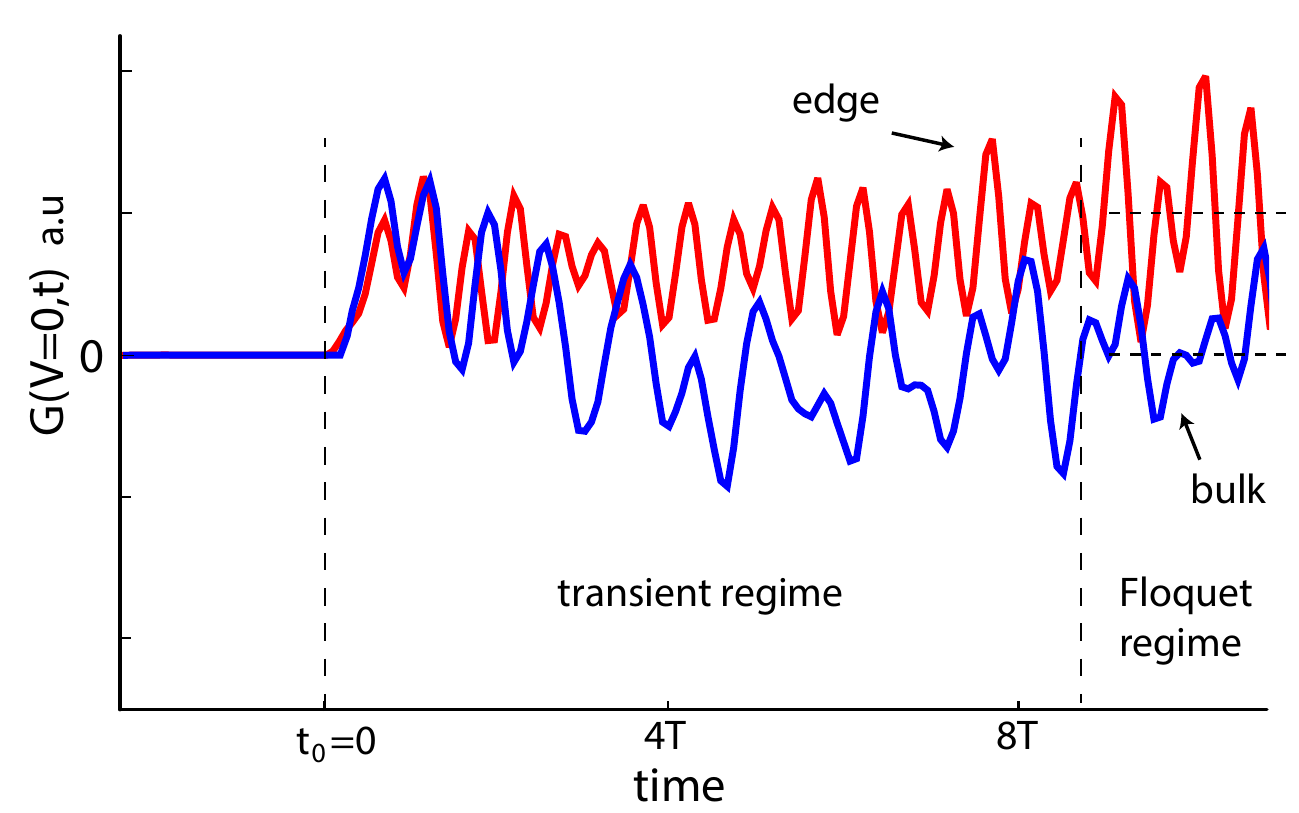}}
\caption{(Color online) Typical time-dependent conductance at zero bias. The field is turned on 
at $t_0$ and the system evolves towards a regime in which the conductance becomes periodic. 
Horizontal dashed lines indicate the time-averaged
component in the Floquet regime. The transient regime lasts about 10-20 periods for our system parameters. 
A background conductance has been subtracted. Note that in the Floquet regime higher harmonics have been generated.
Here we take $N_x=N_y=14$ and rest of the parameters are the same as in Fig.~\ref{fig:conduc}.}
 \label{fig:G0time}
\end{figure}

\textit{Acknowledgments}.
We thank A. Kamenev, F. Juan, A. Lobos, S. Coh, Z. Pedramrazi and J. Velasco for useful discussions. 
B.M.F. acknowledges partial support from Conacyt, J.E.M. from NSF DMR-1206515 and the DARPA MESO program and 
J.P.D. from NWO and DAAD. Computer resources were provided by NERSC under Contract No. DE-AC02-05CH11231.

\appendix

\section{Wideband limit}
\label{sec:wbl}
In the wideband limit (WBL), the coupling to reservoirs are energy independent,
$J_{T k\sigma ,is}=\delta_{\sigma s}\delta_{i T} J_{T}$ and $J_{B k\sigma ,is}=\delta_{\sigma s} J_{B}$,
and Eqn.~\ref{eqn:gen_current} becomes
\begin{align}
I_{T}(t) &= -\frac{e\Gamma_T}{\hbar} \textrm{Im} \big[G^{<}_{TsTs}(t,t)  \label{eqn:current_gen} \\
&\hspace{35pt}+\hspace{-3pt}\int \hspace{-3pt} dt_1 \hspace{-3pt}\int\hspace{-3pt} \frac{d\epsilon}{\pi} e^{i\epsilon(t-t_1)}f_{T}(\epsilon)G^{r}_{TsTs}(t,t_1)\big]. \nn
\end{align}
Summation over repeated spin indices is implied. We can assume that the retarded 
and advanced Green functions of the system do not depend on the chemical 
potential of the reservoirs. Then the differential conductance $dI_{T}(t)/dV$ is 
\begin{align}
G_T(V,t) & = \frac{e^2}{h}\bigg[\pi\Gamma_T^2 \textrm{Re} \hspace{-3pt} \int \hspace{-4pt} dt_1 \hspace{-5pt} \int \hspace{-3pt} dt_2  	G^{r}_{TsTs'}(t,t_1)G^{a}_{Ts'Ts}(t_2,t) & \nonumber \\ 
 & \hspace{110pt}\times \frac{(t_2-t_1)e^{i eV (t_2-t_1)}}{\beta \sinh (\frac{\pi(t_1-t_2)}{\beta})}  &\nonumber \\
 & \hspace{0pt}+ 2\pi\Gamma_T \textrm{Im}\hspace{-3pt}\int \hspace{-3pt} dt_1 G^{r}_{TsTs}(t,t_1)\frac{(t-t_1)e^{i eV (t-t_1)}}{\beta\sinh (\frac{\pi(t_1-t)}{\beta})}\bigg]. & 
\label{eqn:G_gen_time}
\end{align}
We note that at zero temperature and for time independent Hamiltonians, 
Eqn.~\ref{eqn:G_gen_time} gives $G_T(V)= (e^2/h)[M_T- \textrm{Tr}[r_{T}r^{\dagger}_{T}]]$, as
expected, where
 $M_T =2\pi\Gamma_T \rho_T(eV) = -2 \Gamma_T\textrm{Im}\sum_s G^{r}_{TsTs}(eV)$ is the 
number of channels which is proportional to the \textit{local} density of 
states and $(r_{T})_{ss'}=\Gamma_T G^{r}_{TsTs'}(eV)$ is the 
\textit{local} electron reflection matrix.  In Fig.~\ref{fig:G0time},
we see an example of the conductance as a function of time from Eqn.~\ref{eqn:G_gen_time}.

\section{Periodic Hamiltonian}
\label{sec:periodic_H}
Some analytic progress can be made for free electrons that are subject to a periodic drive 
$H_S(t+T)=H_S(t)$ with period $T=2\pi/\Omega$.  All two-time correlators 
become periodic, e.g., $G^{r,a,<}(t+T,t'+T)=G^{r,a,<}(t,t')$ and Eqn.~\ref{eqn:current_gen} becomes
\begin{align}
\bar{I}_T = \frac{e\Gamma_T }{\hbar}\big[ \int \frac{d\epsilon}{2\pi} f_{T}(\epsilon)\bar{A}_{TsTs}(\epsilon) -\textrm{Im} \bar{G}^{<}_{TsTs} \big],
\label{eqn:current_periodic_H}
\end{align}
where $\bar{A}_{TsTs}(\epsilon) = -2\textrm{Im}G^{r}_{TsTs}(0,\epsilon)$ 
is the time-averaged nonequilibrium spectral function and 
we used the Wigner representation for the Green functions.  We also defined,
\begin{align}
\bar{G}_{is,js'}^{<} &= i \sum_{n,\alpha} \int \frac{d\epsilon}{2\pi} f_{\alpha}(\epsilon) G_{is,\ell\sigma}^{r}(n,\epsilon + n\Omega/2) \times \\ 
&\hspace{65pt}\Gamma_{\ell\sigma, \ell'\sigma'}^{\alpha} G_{js',\ell' \sigma'}^{r}(n,\epsilon +n\Omega/2)^{*}. \nn
\end{align}
%
The time-average occupation can be written as $\bar{n}_{is} = -i \bar{G}_{is,is}^{<}$. 
Eqn.~\ref{eqn:current_periodic_H} has an intuitive interpretation\cite{Jauho1994}.
The first term is proportional to the occupation of the tip and 
the \textit{local} sample spectral function and hence can be interpreted as the current flowing into the sample from the tip.
The second term is proportional to the time-averaged local occupation probability at the position of the tip and can be interpreted as the the probability of charge 
flowing out of the sample to the tip. 
Note that since the couplings to the tip and bath are not proportional we cannot in general write the above 
expression in a Laundauer-type form. However the time-averaged 
conductance at zero temperature can be written in terms of general scattering matrices of the central region,
from Eqn.~\ref{eqn:G_gen_time},
\begin{align}
\bar{G}_T(V) = \frac{e^2}{h}\sum_n \{ M_T(n)  \delta_{n,0} - \textrm{Tr}[r_{T}(n)r^{\dagger}_{T}(n)] \},
\label{eqn:conductance_periodic} 
\end{align}
where we defined $M_T(n) = 2\pi \Gamma_T \rho_{T}(n,eV) = -2 \Gamma_T \textrm{Im} G^{r}_{TsTs}(n,eV)$
and $\rho_{T}(n,eV)$ is the 'density of states' for $n$-th band and the electron 
reflection matrix $[r_{T}(n)]_{ss'} = \Gamma_T G^{r}_{TsTs'}(n,eV+n\Omega/2)$.
The first term is a \textit{local} density of states and the second is a contribution from electron tunneling.
For non-interacting electrons we can write  $G^{r}(t,t')$ in terms of the eigenvalues and right and left eigenvectors 
of the Floquet Hamiltonian $\mathcal{H}_F=H_S(t) + \Sigma^{r}- i\partial_t$,
i.e., $\mathcal{H}_F \phi^{R \delta m}= \epsilon_{\delta m} \phi^{R\delta m}$,
$(\phi^{L \delta m})^{\dagger}\mathcal{H}_F = \epsilon_{\delta m} (\phi^{L\delta m})^{\dagger}$,
as  $G^{r}_{is, js'}(n,\omega) = \sum_{\delta m} \phi_{is, n}^{R\delta \bar{m}}  (\phi_{js', 0}^{L\delta \bar{m}})^{*}/(\omega - \epsilon_{\delta m} - n\Omega/2)$,
where the quasienergies are complex, $\epsilon_{\delta m} = \epsilon_{\delta m} - i \gamma_\delta$ and we defined $\bar{m}=-m$. 
Then the occupation of a given site at zero temperature is from Eqn.~\ref{eqn:gr_gless},
\begin{align}
\bar{n}_{i\sigma}& = \hspace{-0pt}\sum_{\alpha,n}\hspace{-0pt} \sum_{\substack{\ell\sigma\\ \ell'\sigma'}} \hspace{-0pt}\sum_{\substack{\delta m\\ \delta' m'}} \hspace{-0pt} \Gamma^{\alpha}_{\ell\sigma,\ell' \sigma'} \frac{\phi^{R\delta m}_{i\sigma,n} (\phi^{L\delta m}_{\ell\sigma,0})^{*} (\phi^{R\delta' m'}_{i\sigma,n})^{*} \phi^{L\delta' m'}_{\ell'\sigma',0}}{i\epsilon_{\delta m} - i\epsilon^{*}_{\delta' m'}} \nn\\ 
&\hspace{100pt}\times\bigg[1 - \frac{1}{2\pi i}\log\bigg(\frac{\epsilon_{\delta m}}{\epsilon^{*}_{\delta' m'}}\bigg) \bigg].
\end{align}
and the electron-reflexion matrix becomes,
\begin{align}
[r_{T}(n,eV)]_{ss'} = \Gamma_T \sum_{\delta m} \frac{\phi_{Ts, n}^{R\delta \bar{m}} (\phi_{Ts', 0}^{L\delta \bar{m}})^{*}}{eV - \epsilon_{\delta m}}.
\end{align}

\bibliographystyle{apsrev}

\end{document}